# PLL Based Sub-/Super-synchronous Resonance Damping Controller for D-PMSG Wind Farm Integrated Power Systems

Songhao Yang, *Member, IEEE*, Ruixin Shen, *Student Member, IEEE*, Jin Shu, *Member, IEEE*, Tao Zhang, *Student Member, IEEE*, Yujun Li, *Member, IEEE*, Baohui Zhang, *Fellow, IEEE*, Zhiguo Hao, *Member, IEEE*

*Abstract*—Existing sub-/super-synchronous (SSO) suppression methods for the direct-drive permanent magnet synchronous generators (D-PMSG) integrated power systems are mainly achieved by external devices or sub-synchronous resonance damping controller (SSRDC) at the converters, facing challenges of considerable control costs, complex parameters tuning, or inadaptability to various operating conditions. To address these problems, this paper proposes an adaptive SSRDC based on the phase-locked loop (PLL) for D-PMSG integrated power systems. Firstly, the PLL parameter is found critical to SSO suppression by a comprehensive sensitivity analysis on the dominant poles of the impedance closed-loop transfer function. Motivated by this finding, this paper then designs a PLL-based SSRDC, which features a simple structure, easy parameter tuning, and flexible adaptability to various operating modes. The simplicity in structure is guaranteed by the avoidance of phase compensation. Benefiting from the simple structure, only one key parameter needs to be tuned. Moreover, two principles of parameter tuning are proposed to enhance the efficiency, robustness, and adaptability of the proposed SSRDC. The controller-hardware-in-the-loop (CHIL) tests verify the validity of the proposed SSRDC under various operating conditions. Finally, some concerns about this method such as frequency estimation, computational efficiency and potential impacts on PLL are thoroughly analyzed and clarified.

## Nomenclature

| | |
|---|---|
| $V_1$ | Magnitude of the integrated voltage |
| $V_{dc}$ | Reference voltage of the bus |
| $k_m$ | Gain of the modulator |
| $k_d, k_f$ | Decoupling, feedforward coefficient |
| $k_p$ | Proportional gain of the PLL |
| $k_i$ | Integral gain of the PLL |
| $k_{ip}$ | Proportional gain of the inner current loop |
| $k_{ii}$ | Integral gain of the inner current loop |
| $f_{sso}$ | Difference between the super-synchronous frequency and the fundamental frequency |
| $\omega_{sso}$ | Center frequency of the band-pass filter |
| $H_0$ | Gain of the band-pass filter |
| $\zeta$ | Damping coefficient of the band-pass filter |
| $v_d, v_q$ | D/q axis of the grid side voltage |
| $k_{sso}$ | Tuned gain of the damping controller |
| $X_l$ | Reactance of the transmission line |
| $X_T$ | Reactance of the transformer |
| $R_l$ | Resistance of the transmission line |
| $n$ | Number of online wind turbines |
| $t_e$ | Time delay of sub/super-synchronous frequency estimation |
| $Z_{pmsg}$ | Impedance of wind turbine |
| $Z_{PMSG}$ | Impedance of wind farm |

## I. Introduction

### A. Motivation

RECENTLY, renewable energy power generation led by wind power has received widespread attention and utilization. Among them, the direct-drive permanent magnet synchronous generators (D-PMSG) have an appealing prospect due to the high energy conversion efficiency and low maintenance cost [1, 2]. However, the integration of the large-scale D-PMSG wind farms into power systems brings new challenges. Several sub-/super-synchronous oscillation (SSO) accidents have occurred in practical power systems such as in Guyuan of North China [3] and Hami of Northwest China [4]. These SSO accidents, caused by the integration of D-PMSG wind farms, finally led to the tripping of wind farms and shaft torsional vibration of nearby thermal power units, seriously threatening the stable operation of the power system. Thus, an effective SSO suppression scheme for D-PMSG wind farm integrated power systems is urgently demanded.

### B. Related Works

Currently, the research on SSO suppression is mainly focused on the doubly-fed induction generator (DFIG) integrated systems, and few of them are dedicated to the D-PMSG integrated systems. Due to the similarities in the converter control and grid integration topology of wind turbines,

This work was partially supported by National Natural Science Foundation of China (52007143), China Postdoctoral Science Foundation (2021M692526) and Open Fund of State Key Laboratory of Operation and Control of Renewable Energy & Storage Systems (China Electric Power Research Institute).
Songhao Yang, Ruixin Shen, Tao Zhang, Yujun Li, Baohui Zhang, Zhiguo Hao are with Shaanxi Key Laboratory of Smart Grid, Xi'an Jiaotong University, Xi'an, China (e-mail: {songhaoyang, zhghao @xjtu.edu.cn,).
Jin Shu is with Xi'an Thermal Power Research Institute CO.LTD.

some of the SSO suppression methods designed for the DFIG systems are of great reference for the D-PMSG systems. These related methods can be categorized into three kinds, namely 1) the beforehand measures in the system planning and operation stage, 2) the emergency measures after serious SSO occurrence and 3) the active damping control.

The first category of methods attempts to avoid the SSOs in the system planning and operation stage by increasing the grid strength and reducing incentives for SSOs [5]. In the planning stage, enhancing the grid connection[6] and redesigning the coordination between control loops[7, 8] are beneficial to avoiding the SSO. However, these measures are only effective for unbuilt power systems. For an already existing system, the operation mode adjustment can also alleviate the occurrence of SSO. Switching wind turbines selectively [9] and reactive power regulation[10] have been proven effective in practice. However, the occurrence of SSO cannot be fundamentally avoided by such measures due to the limited margin of operation adjustment[11]. The second category of methods, namely the emergency protection, trip the oscillated wind turbines by the sub-synchronous frequency relay[12]. These measures are usually considered as the last resort to protect the system in consideration of the serious consequences.

The active damping control methods, converging the SSO by providing additional damping, have attracted wide attention. The sub-synchronous resonance damping controllers (SSRDC) can be achieved by external devices or wind turbine converters. The existing external devices, such as the flexible AC transmission system (FACTS) devices and high voltage DC (HVDC) converters, can be an alternative for SSO mitigation if they are equipped with additional damping controllers. SSRDCs equipped with the Thyristor Controlled Series Capacitor (TCSC)[13], Gate-Controlled Series Capacitors (GCSC)[14], Static Synchronous Compensator (STACOM)[15, 16] can mitigate SSO effectively by changing the system impedance and moving the eigenvalues of the system to the left half-plane[17]. However, primarily designed for power flow regulation, these devices cannot be always available for SSO suppression. To mitigate SSO more generally, a dedicated device was designed in [18]. The proposed method can provide active damping by injecting sub-synchronous frequency currents. Literature [19] further improved its adaptability by online capturing the sub-synchronous frequency with a sub-synchronous frequency estimator. Analysis shows that the marginal cost of such plant-level SSO suppression schemes would decrease with the increase in wind farm scale[19]. Therefore, these special-purpose devices could be a relatively competitive solution despite the additional investment.

The SSRDC on wind turbine converters (WTCs) may be one of the most concerning methods because it can be achieved in a software manner without hardware modification. The WTCs can provide damping by adjusting the control parameters or modifying the structure. Literature [20] adjusted the proportional parameter of the rotor side converter (RSC) of DFIG to prevent the SSO. In [21], the SSRDC implemented in converters of DFIGs can provide damping at only the frequency of interest by modifying the reference current of the inner current controller loop. However, the actual effects of these methods may not be good as expected due to the limited adjustment space of parameters. New controller structures of converters are also proposed to avoid the SSO's occurrence. The PI controller of the grid side converter (GSC) is replaced by a nonlinear controller in [22], and the inner current loop at RSC is replaced by a direct controller in [23]. However, these nonlinear controllers feature complicated design and complex parameter tuning, thus they are rarely implemented in practice[24]. Compared with the above two kinds of methods, adding a damping control loop to existing converters is much easier to implement [25-28]. Literature [25] designed an SSRDC at the outer loop of GSC and established a parameters look-up table. By adjusting the parameters accordingly, the SSRDC can provide effective damping for various operating conditions. However, constructing such a complete lookup table is time-consuming and challenging. Since the mechanism of SSO caused by DFIG and PMSG is not identical[26], literature [27] analyzed the key factors of PGMG-induced SSOs and designed an SSRDC at the inner current loop of GSC. However, it still faces the problem of complicated parameter tuning. To this end, literature [28] proposed an automatically parameter-tune method for SSRDC based on the improved particle swarm optimization algorithm.

*C. Main Contribution*

An SSRDC that is simple tuning and adapted to various conditions is urgently needed for D-PMSG integrated power systems. Most of the existing SSRDCs are designed on the converters, but rare of them consider the influence of the phase-locked loop (PLL) though it has been found to be essential to the stability of the D-PMSG integrated system[29]. This paper investigates the SSO suppression method based on PLL. The contributions include:

1) The essential impacts of PLL bandwidth on system stability are demonstrated based on the sensitivity analysis of impedance closed-loop transfer function's dominant poles. This finding motivates the design of the PLL-based SSRDC.

2) A PLL-based SSRDC is proposed. The SSRDC has the advantages of simple structure, easy parameter tuning, and strong adaptability to various operating conditions. The simple structure is guaranteed by the avoidance of phase compensation. Benefiting from the simple structure, only one key parameter needs to be tuned. To balance the simplicity, robustness, and adaptability of the SSRDC, two principles of parameter tuning are proposed, namely pre-determining parameters under current worst conditions and updating parameters online when necessary.

3) A Controller-hardware-in-the-loop (CHIL) test is carried out to verify the efficiency and adaptability of the proposed SSRDC under various operating conditions.

4) Some concerns about the proposed SSRDC are thoroughly discussed and clarified. This method has good tolerance to the impedance model and frequency estimation algorithm. The SSRDC has high computation efficiency of parameter tuning, and it does not influence the normal steady-state and dynamic response of PLLs.

The rest of the paper is organized as follows. Section II analyzes the essential impacts of PLL parameters on system stability. Section III presents the details of the PLL-based

SSRDC. Through the CHIL test in section IV, the effectiveness and adaptability of the proposed SSRDC are verified. Section V discusses and clarifies the concerns about the proposed method. Finally, conclusions are drawn in Section VI.

## II. SSO Analysis of D-PMSG Wind Farm Integrated Power Systems

### A. SSO Analysis Method Based on Impedance Closed-loop Transfer Function's poles

In existing research, impedance-based SSO analysis methods are becoming popular for their simple and independent modeling[30]. This paper adopts the idea of impedance modeling and analyzes the SSO stability based on the dominant poles (DP) of the impedance closed-loop transfer function.

Fig. 1 presents the small-signal equivalent circuit of the D-PMSG wind farm integrated power system. The wind farm is equivalent to a Norton Circuit consisting of an ideal current source $I_s(s)$ and the external equivalent impedance of wind farm $Z_{PMSG}(s)$. AC power grid is equivalent to a Thevenin Circuit consisting of an ideal voltage source $U_g(s)$ and the grid impedance $Z_g(s)$. $I_g(s)$ and $V_{PCC}(s)$ denote the grid-integrated current and voltage. $Z_{PMSG}(s)$ adopts the impedance model proposed in [31] for its simplicity and generality, and $Z_g(s)$ is determined in advance or measured online.

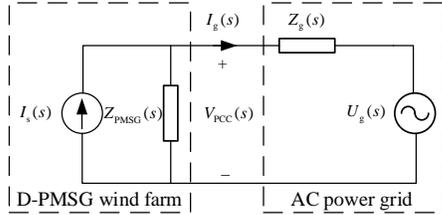

Fig. 1. Small-signal representation of equivalent circuit diagram of D-PMSG integrated power systems.

Since $I_g(s)$ is the interaction current between wind farm and grid, its state can reflect the stability of the wind farm integrated power system. $I_g(s)$ can be obtained as:

$$I_g(s) = I_s(s) \cdot \frac{1}{1+Z_g(s)/Z_{PMSG}(s)} - U_g(s) \cdot \frac{1}{Z_g(s)+Z_{PMSG}(s)}, \quad (1)$$
$$= I_s(s) \cdot G_1(s) - U_g(s) \cdot G_2(s)$$

where $G_1(s)=1/(1+Z_g(s)/Z_{PMSG}(s))$, $G_2(s)=1/(Z_g(s)+Z_{PMSG}(s))$.

According to the superposition theorem, the circuit in Fig.1 is equivalent to the superposition of two sub-graphs in Fig. 2.

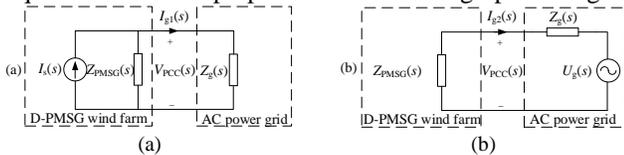

Fig. 2. Small-signal representation of D-PMSG wind farm systems considering (a) wind farm current disturbances. (b) grid voltage disturbances.

Since $I_s(s)$ and $U_g(s)$ are linearly independent, the system stability is determined by $G_1(s)$ and $G_2(s)$ together. Comparing the formula of $G_1(s)$ and $G_2(s)$, it is easy to find that

$$P(G_1(s)) \subseteq P(G_2(s)), \quad (2)$$

where $P(G(s))$ is the pole set of $G(s)$. For a linear system, its stability is determined by its DP. The real part of DP reflects the system's damping level, a negative value indicating the system is stable. And the imaginary part of DP reflects the oscillation frequency. Therefore, the SSO analysis of the D-PMSG wind farm integrated power system can be carried out based on $G_2(s)$'s DP.

To simplify the analysis, remark $G_2(s)$ as $G(s)$:

$$G(s) = \frac{1}{Z_g(s)+Z_{PMSG}(s)}. \quad (3)$$

### B. Influence Factors Analysis of SSO

Based on the above analysis, the key factors of SSO in the grid integrated D-PMSG power system can be analyzed through their sensitivity to $G_2(s)$'s DPs. Fig.3 presents the control structure of the converter. The following factors are discussed.

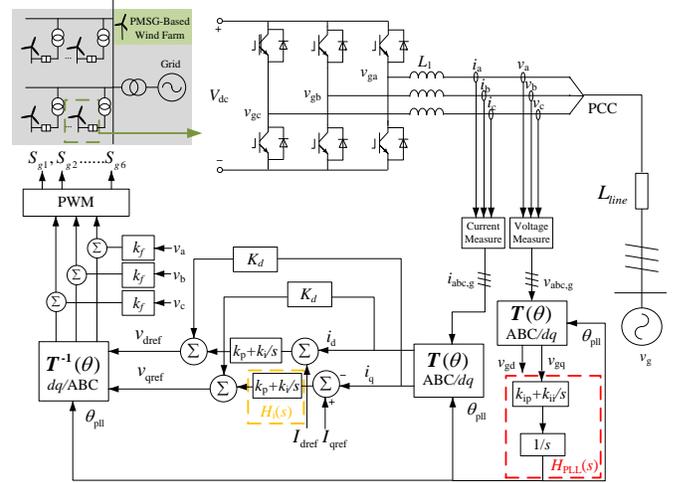

Fig. 3. The control structure of the PMSG converter

1) The parameters of grid side converter (GSC), i.e. the PI parameters of the PLL ($k_p$, $k_i$), and the inner current loop ($k_{ip}$, $k_{ii}$) are fully discussed as they mainly determine the dynamics of D-PMSG. Note that the direct-voltage control loop is not considered in the impedance modeling because its influence on SSOs caused by PMSGs is negligible [8].

2) The operating parameters are also discussed, including the output current of each D-PMSG unit($I_1$), number of online units($n$), and grid-side impedance ($R$, $X$).

To qualitatively analyze the influence of the above factors on the DPs, the idea of sensitivity analysis is adopted. Denote DP as

$$DP = A + B \cdot i, \quad (4)$$

where $A = \text{Re}\{DP\}$, $B = \text{Im}\{DP\}$.

Thus, the sensitivity of parameters to system stability can be written as their sensitivity to the real part of DP.

$$S_x = \frac{\partial A}{\partial x}, \quad (5)$$

where $x$ denotes the above-mentioned parameters.

The impedance of one PMSG turbine is computed using the model proposed in [31].

$$Z_{pmsg}(s) = \{k_m V_{dc}[H_i(s-j2\pi f_1) - jK_d] + sL_1\} \cdot$$
$$\left\{1 - k_m V_{dc} k_f - \left[\frac{T_1}{2}e^{j\varphi_{f_1}} + H_i(s-j2\pi f_1)\frac{I_1}{2}e^{-j\varphi_{f_1}}\right.\right. \quad (6)$$
$$\left.\left. -jK_d \frac{I_1}{2}e^{-j\varphi_{f_1}}\right] \times T_{PLL}(s-j2\pi f_1)\frac{k_m V_{dc}}{V_1}\right\}^{-1}$$

where $V_1$ is the magnitude of the integrated voltage, $k_m$ denotes the gain of the modulator, $k_d$, and $k_f$ denote the decoupling coefficient and feedforward coefficient respectively. $V_{dc}$ denotes the reference voltage of the bus voltage. $\varphi_{i1}$ denotes the phase current of $I_1$, $H_i(s)=k_{ip}+k_{ii}/s$, $H_{PLL}(s)=(k_p+k_i/s)/s$, $T_{PLL}(s)=V_1H_{PLL}(s)/[1+V_1H_{PLL}(s)]$. $T_1$ and $\varphi_{T1}$ are reversely computed by:

$$\frac{T_1}{2}e^{\pm j\varphi_{T1}} = \frac{j2\pi f_1 L_1 + V_1(1-k_m V_{dc}k_f)}{k_m V_{dc}}. \quad (7)$$

To simplify the analysis, the assumption of uniform parameter wind turbines is adopted, that is, the control parameters of the wind turbines are all the same. Therefore, the impedance of the wind farm can be simplified as

$$Z_{PMSG} = [Z_{pmsg}(s) + X_T(s)]/n, \quad (8)$$

where $X_T$ denotes the impedance of the box transformer, and $n$ is the number of wind turbines.

Theoretically, through (3)-(7), the analytical formulas of $S_x$ can be obtained for different parameters. However, due to the high order and complexity of $Z_{PMSG}$, it is difficult to get $S_x$ analytically. Instead, $S_x$ can be calculated numerically by (9).

$$S_x = \lim_{\Delta x \to 0} \frac{A(x+\Delta x) - A(x)}{\Delta x} \quad (9)$$

Varying the value of each parameter in the range of [0.5,1.5] p.u., $S_x$ is obtained accordingly, as shown in Fig. 4.

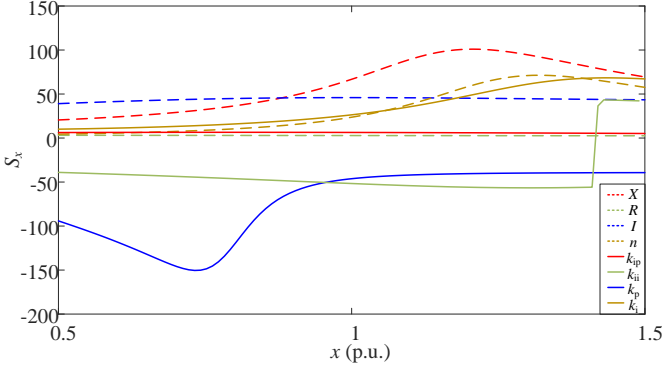

Fig. 4. Sensitivity analysis of each parameter to system stability.

According to (5), if $S_x$ is negative, DP moves left on the s-plane with the increase of parameter $x$, indicating the system stability is enhanced. On the contrary, if $S_x$ is positive, the increase of this parameter will deteriorate the stability of the system. Moreover, a greater absolute value of $S_x$ implies the system stability is more sensitive to the parameter changes.

Fig. 4 presents the sensitivity index of each parameter to system stability. The dashed lines show that the increase of operating parameters including $X$, $I$, $n$, and $R$ have negative effects on system stability. In other words, SSO more easily occurs in conditions of larger grid reactance, more online wind turbines, and larger output power. Conversely, the increase in the control parameters represented by solid lines can enhance the system's stability.

For current inner loop parameters, $k_{ip}$ is not recommended for SSO suppression as $S_{k_{ip}}$ is nearly zero, indicating that system stability is not sensitive to this parameter. Though $k_{ii}$ has a satisfactory absolute value, it is not suitable for SSO suppression because the sign of $S_{k_{ii}}$ is not constant. When $k_{ii}$ increases, the system stability is firstly enhanced, but then deteriorated. The uncertainty of parameter impacts will increase the difficulty of SSO suppression design. As for the PLL parameters, $k_i$, and $k_p$, both have the potential for SSO suppression. $k_p$ is more recommended as its sensitivity is very high, and more importantly, the high sensitivity is global when compared with $k_i$. Therefore, increasing $k_p$ to stabilize the system always has a significant and ideal effect.

To conclude,

1) PLL is essential to the stability of the D-PMSG wind farm integrated power systems. The system tends to be more stable by increasing the $k_p$ of PLL, and this method has globally high sensitivity. Inspired by this finding, this paper designs a novel SSRDC at the PLL.

2) The worst condition that causes SSO is revealed, which can be used for SSRDC parameter pre-tuning and case verification.

### III. DESIGN OF PLL BASED SSRDC

#### A. Design Principle

An SSRDC with simple parameter tuning and flexible adaptability to various operating conditions is required. Therefore, the following design principles should be met.

1) *Simple structure*: To meet the needs of easy tuning, the structure of SSRDC should be furtherly simplified. Generally, an SSRDC contains a bandpass filter (BPF), phase compensation, gain, and limiter. Among them, time constants $T_1$, $T_2$, and order $n$ in the phase compensation loop need to be tuned to coordinate with other loops under different operating modes, which significantly increases the complexity of parameter tuning. If the parameters tuning of the phase compensation can be avoided, the tuning process will be significantly simplified.

2) *Flexible adaptability:* To achieve SSRDC's flexible adaptability, the SSRDC is expected to suppress SSO quickly under various operating conditions. Besides, SSRDC should not start up under normal conditions as the control parameters of the D-PMSG controller are optimal by default.

3) *Robustness:* To ensure the suppression effect, enough stability margin must be reserved in the designed SSRDC. A simple solution is to tune the parameters under the worst condition.

#### B. SSRDC Design

Based on the above principles, this paper designs a novel SSRDC at the PLL. The schematic diagram is shown in Fig. 5

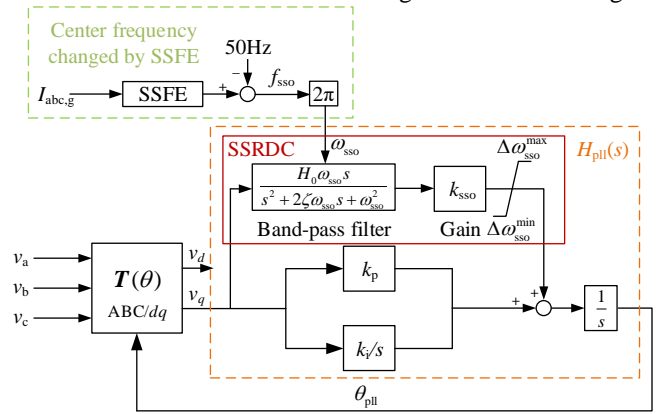

Fig. 5. The proposed PLL-based SSRDC.

The SSRDC proposed in this paper is composed of a BPF with a changeable center frequency, super-synchronous frequency estimator (SSFE), gain, and a limiter. The SSFE is assumed to accurately measure the super-synchronous frequency component of the wind farm's output current with a time delay in the range of 100ms-500ms [19, 32]. The center frequency of BPF defaults to 0, and is updated to the difference between the super-synchronous frequency and the standard frequency once SSO occurs. The gain $k_{sso}$, is the only parameter to be tuned.

The basic working logic of the proposed SSRDC is given as follows.

1) Under normal operating conditions, the super-synchronous frequency captured by the SSFE is 0 Hz. Whatever $v_q$ inputs, the output of SSRDC is always 0. In other words, the SSRDC does not affect the dynamics of D-PMSGs under normal operating conditions.

2) Once SSO occurs, $v_q$ consists of a DC component and AC components whose frequencies are the differences between the sub/super-synchronous frequency and fundamental frequency. Then the AC components pass through the BPF and increase the $k_p$ of the PLL. The transfer function of the PLL based SSRDC then changes to

$$H_{pll}(s) = \left[ k_p + \frac{k_i}{s} + \left( \frac{H_0 \omega_{sso} s}{s^2 + 2\zeta\omega_{sso}s + \omega_{sso}^2} \right) \cdot k_{sso} \right] \cdot \frac{1}{s}, \quad (10)$$

where $k_p$ and $k_i$ are the original PI parameters of the PLL, $k_{sso}$ is the gain of the SSRDC, $\omega_{sso}$ is the angular velocity of the super-synchronous current captured by the $dq$ axis, $H_0$ is the gain of the BPF, and $\zeta$ is the damping coefficient of the BPF.

When SSO occurs, the oscillation frequency of D-PMSG is super-synchronous according to [33], and a sub-synchronous frequency component is coupled[30]. Therefore, if $f_p$ denotes the super-synchronous frequency, the coupled sub-synchronous frequency will be $2f_1-f_p$, where $f_1$ is the fundamental frequency of the power grid. After the Park's transformation, the coupled frequencies change from $f_p$ and $2f_1-f_p$ to $f_p-f_1$ and $f_1-f_p$, respectively.

Denote $\omega_{sso1}=2*\pi*(f_p-f_1)$ and $\omega_{sso2}=2*\pi*(f_1-f_p)$, thus

$$\omega_{sso} = \omega_{sso1} = -\omega_{sso2}. \quad (11)$$

Note that the frequency captured by SSFE is the super-synchronous component, namely $\omega_{sso1}$. When the AC components pass through the BPF, the frequency response of the BPF to the sub/super-synchronous components become

$$H_{BPF}(j\omega_{sso1}) = H_{BPF}(j\omega_{sso2}) = \frac{H_0}{2\zeta} \quad (12)$$

Thus, the respective phase errors produced by the BPF are

$$\Delta\varphi_1 = \Delta\varphi_2 = 0. \quad (13)$$

Furthermore, the transfer function of PLL can be further simplified as:

$$\begin{cases} H_{pll\_sso}(s) = \left( k_p + \frac{k_i}{s} + \frac{H_0}{2\zeta} \cdot k_{sso} \right) \cdot \frac{1}{s} = \left( k_p' + \frac{k_i}{s} \right) \cdot \frac{1}{s} \\ H_{pll\_fun}(s) = \left( k_p + \frac{k_i}{s} \right) \cdot \frac{1}{s} \end{cases}, \quad (14)$$

where $k_p' = k_p + H_0 \cdot k_{sso}/2\zeta$.

$H_{pll\_sso}(s)$ and $H_{pll\_fun}(s)$ are the transfer function of PLL for SSO components and non-SSO components, respectively. It shows that the proposed SSRDC only provides damping for the SSO components, and non-SSO components are not affected. Moreover, the PLL bandwidth changes for SSO components only when an SSO event occurs. After the risk of SSO is confirmed to be eliminated, $\omega_{sso}$ is reset to 0, and therefore SSRDC is de-activated. Besides, the proposed SSRDC does not require phase compensation compared with the existing SSRDC. This simple structure is achieved by SSFE-based center frequency changeable BPF. This feature of the SSRDC significantly releases the parameter tuning task and enhances its practicability.

### C. Online Parameter Tuning

The parameters that need to be tuned in (14) consist of $k_{sso}$, $H_0$, and $\zeta$. As $H_0$ and $k_{sso}$ are an entirety, $H_0$ can be set as 1 for simplification. And the value of $\zeta$ is determined following the classic second-order BPF[34]. Therefore, only $k_{sso}$ of SSRDC needs to be tuned.

As mentioned above, the value of $k_{sso}$ should meet the requirements of simplicity, robustness, and adaptability. However, these requirements are conflicted to a certain extent. For example, to make it simple, it is ideal that the number of parameter tuning is the lowest, and the computation time for each tuning is the shortest. But the robustness of SSRDC requires that enough stability margin should be reserved, thus the value of $k_{sso}$ should be large enough, indicating a longer computation time for parameter tuning. Moreover, the adaptability of SSRDC requires the parameters updated frequently, which is clearly in conflict with the principle of simplicity.

To achieve a proper balance among these requirements, this paper proposes the following guidelines in parameter tuning.

1) Parameter tuning is pre-performed under the current worst condition.

To ensure that SSRDC is valid under the current operating conditions and enough stable margin is reserved, it is suggested to pre-tune the parameter under the current worst condition. The above analysis has found that the D-PMSG integrated power system is prone to cause SSO under the worst condition that are the greatest grid side impedance, most online units, and maximum unit's output. In this paper, the current worst condition is defined as the number of online units and grid side impedance maintaining the current state, but the online wind turbines output at their rated maximum power. The principle of parameter tuning according to the current worst condition is reasonable as the output of wind turbine changes frequently and randomly in the timescale of minutes, but the number of online units and grid side impedance changes in a long time-scale of hours or days. Besides, as the default parameters of the unit are assumed to be optimal, the value of $k_{sso}$ should be as small as possible. To avoid potential concerns about the performance of the PLL, the upper limit of $k_p'$ is also considered [35, 36].

Based on the above analysis, the algorithm of parameter tuning is given in Table I. It should be noted that if $k_{sso}$ is pre-tuned to 0, it indicates that there's no risk of SSO even under the current worst condition, which can be used to reinitialize the SSRDC.

2) Update the parameter online when necessary.

Although infrequent, the number of online units and the grid side impedance may change significantly under certain situations. For example, the online units will decrease when the wind speed is extremely low or high, and the grid impedance also changes when a fault occurs or the operating mode changes. The number of online units is easy to obtain from the wind farm, and grid side impedance can be measured online by Quadratic Reside Binary (QRB) Sequence method [37]. In such cases, the parameters of SSRDC should be updated online to adapt to the changes in operating conditions.

TABLE I
PARAMETER TUNING PROCESS OF SSRDC

| Goal | SSRDC parameter tuning |
|---|---|
| **Require:** | D-PMSG-based wind farm parameters ($k_p$, $k_i$, $k_{ip}$, $k_{ii}$, $I_{max}$, $n$); grid-side parameters ($U_g$, $X_l$, $X_T$); band-pass-filter parameters ($H_0$, $\zeta$); upper limit of $k'_p$ ($k_{max}$); iterative step size $\varepsilon$. |
| **Ensure:** | the damping ratio of the closed-loop power system $\xi$ leaves a certain margin $\mu_{min}$; min $k_{sso}$. |

1: initialize $\xi = 0$, $k_{sso} = 0$, $i = 1$, $N_{max}=(k_{max}-k_p)/\varepsilon$, $k'_p = k_p$;
2: **while** ($i \leq N_{max}$)
3: calculate dominate pole(DP): DP = $\sigma$+j$\omega$ by solving (3)
4: $\xi_{DP} = -\sigma/(\sigma^2+\omega^2)^{1/2}$;
5: **if** ($\xi_{DP} \leq \mu_{min}$)
6:   $k'_p = k'_p + \varepsilon$; $i = i + 1$;
7: **else**
8:   **break**;
9: **end**
10: **end**
11: update SSRDC gain: $k_{sso} = 2\zeta( k'_p - k_p)/H_0$;

***Recommend***: $H_0$, $k_{max}$, $\mu_{min}$, $\zeta$, and $\varepsilon$ are set to $H_0 = 1$, $k_{max} =0.4$, $\mu_{min} = 0.01$, $\zeta=0.3$, and $\varepsilon = 0.01$ respectively.

The online parameter update is easy to achieve. Firstly, monitor the changes of online units $n$ and grid impedance $X$. Then update $n$ and $X$ and tune the parameter of SSRDC following the algorithm of Table I.

In conclusion, compared with traditional SSRDC, the proposed SSRDC has the following features:

1) The SSRDC features a simple structure and easy tuning attributed to the avoidance of phase compensation.

2) The SSRDC reserves sufficient stability margin by pre-tuning parameters under the current worst condition.

3) The SSRDC ensures its strong adaptability by pre-determining parameters under the current worst condition and updating parameters online when necessary.

4) The SSRDC only needs to update the unit controller parameters online to suppress the SSO. For already installed units, this software-based upgrade is very easy to implement. However, for very old units that cannot change the controller parameters online, the hardware-based upgrade may be a more practical solution [38].

IV. EXPERIMENTAL VERIFICATION

A. *Controller-Hardware-in-the-Loop Platform*

Due to the complexity and encapsulation of the actual wind turbine controllers, it is difficult to simulate the real dynamic response of the wind turbines through simulation models. Therefore, Controller-Hardware-in-the-Loop (CHIL) real-time simulation becomes a convincing way for dynamic analysis of wind turbine integrated power systems[39]. To fully verify the effectiveness of the proposed method, a D-PMSG CHIL test platform is built. The hardware devices include the D-PMSG controller, a real-time digital simulator (RTDS), and two PCs. The D-PMSG controller (Model: NES5412-6400) is produced by NARI TECH, an important power system control and protection equipment manufacturer in China. The proposed SSRDC is embedded in the D-PMSG controller. Besides, RSCAD and MATLAB are used for simulation and parameter tuning, respectively.

Fig. 6 shows the connection of the D-PMSG CHIL test platform. The wind farm integrated power system model is built in the RSCAD of PC1. PC2 connects to the D-PMSG controller through local area network (LAN) 2, which is used for tuning and updating the controller parameters. The RTDS connects PC1 through LAN 1 and connects the D-PMSG controller through D/A and A/D interfaces.

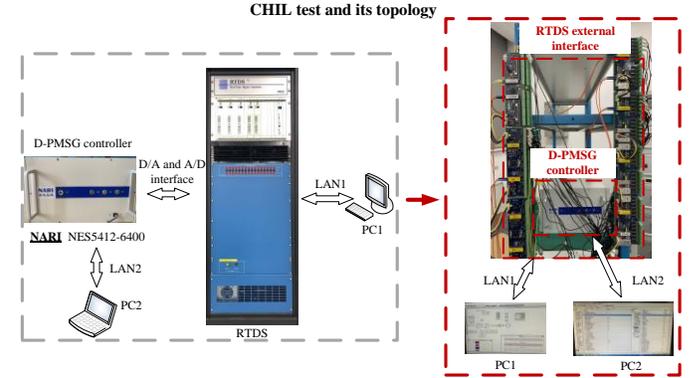

Fig. 6. The connection of the CHIL test platform.

TABLE II
PARAMETERS OF D-PMSG WIND FARM SIDE

| | Symbol | Quantity | Value |
|---|---|---|---|
| Operating parameters | $S_B$ | base capacity | 20MW |
| | $V_s$ | grid side voltage | 690V |
| | $I$ | the output current of each unit | 1650A |
| | $I_{max}$ | the maximum output current of each unit | 2200A |
| | $n$ | number of online units | 6 |
| Control parameters | $k_p$ | the proportional gain of the PLL | 0.09 |
| | $k_i$ | the integrational gain of the PLL | 32 |
| | $k_{ip}$ | the proportional gain of the inner current loop | 0.25 |
| | $k_{ii}$ | the integrational gain of the inner current loop | 120 |
| | $L$ | filter inductance | 0.075mH |

TABLE III
PARAMETERS OF GRID-SIDE

| Symbol | Quantity | Value |
|---|---|---|
| $X_T$ | reactance of transformer | 0.1512p.u. |
| $X_L$ | reactance of transmission line | 0.6553p.u. |
| $R_L$ | resistance of transmission line | 0.0504p.u. |

Fig. 7 presents the information flow of the CHIL test platform. RSCAD is used to build wind speed models, wind turbine models, and system-side component models, which simulate the dynamics of the D-PMSG integrated power system. Then the simulation information is loaded to RTDS through LAN1. The interaction between RTDS and D-PMSG controller is mainly realized by Giga-Transceiver Analog Output (GTAO),

Giga-Transceiver Digital Input (GTDI), and Giga-Transceiver Digital Output (GTDO). The GTAO port is used to output the state variables such as voltage and current (red color signals in Fig. 7) to the controller in an analog form. Reversely, the Pulse Width Modulation (PWM) control signals (purple color signals in Fig. 7) of the D-PMSG controller input to the RTDS through the GTDI port. These controller signals feedback to the RSCAD model to drive the action of the converter and realize the control of D-PMSGs. The GTDO transforms the switch signals such as circuit breaker opening and closing status (brown color signals in Fig.7) from the RSCAD to the controller.

It should be noted that the NARI controller adopts the decoupled double synchronous reference frame phase-locked loop (DDSRF-PLL), which can effectively track the positive sequence component even under unbalanced voltage conditions [40]. Since the voltage waveform is symmetrical when SSO occurs, the negative sequence PI link does not work. In other words, DDSRF-PLL is equivalent to the SRF-PLL in SSO analysis. Therefore, the proposed method can be directly applied to DDSRF-PLL-based D-PMSGs.

Parameters of the D-PMSG wind farm and the grid side are given in Tables II and III, respectively. To reserve enough time margin for frequency estimation, the time delay of SSFE is assumed to be 200ms.

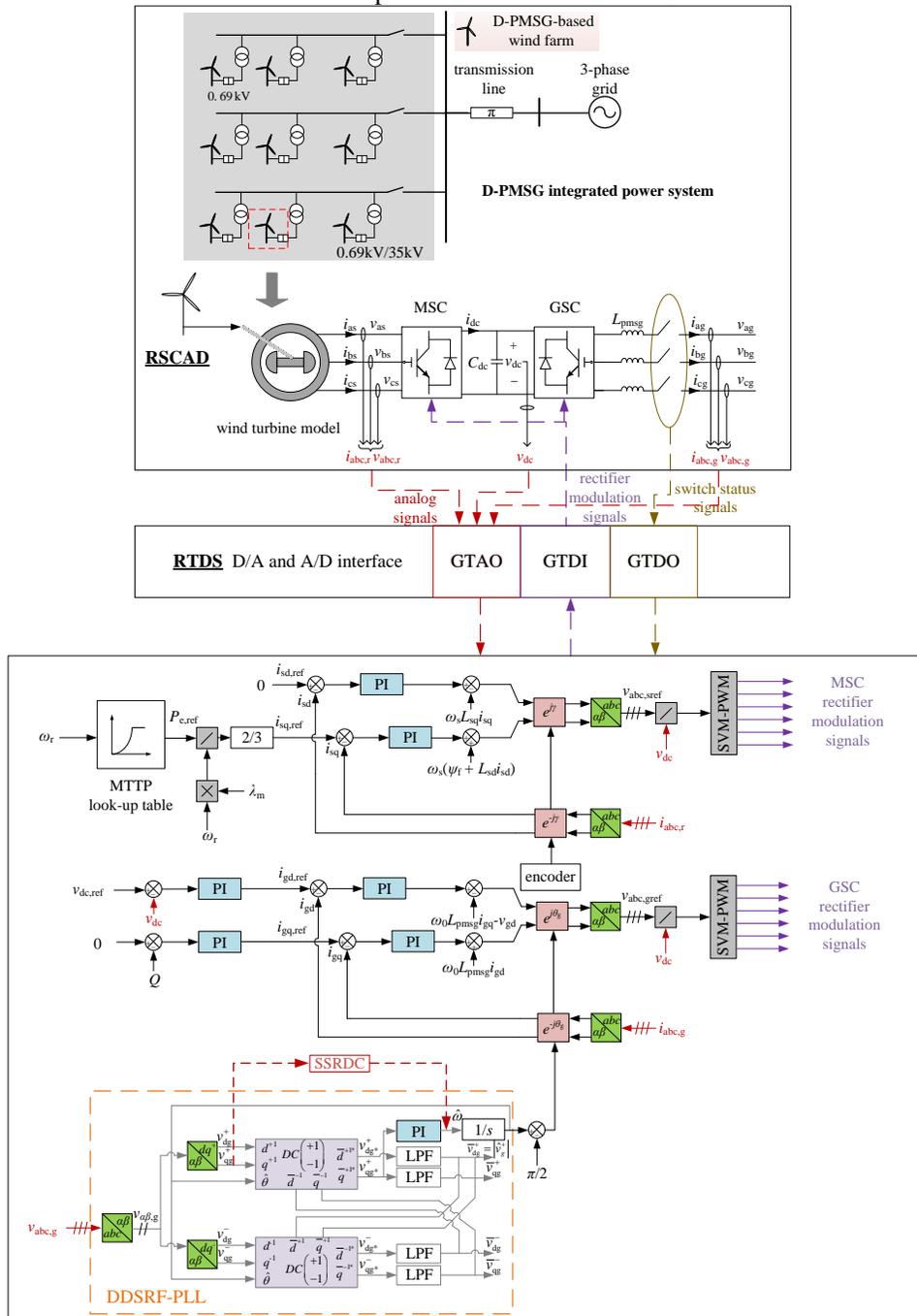

Fig. 7. Information flow of the CHIL test platform

## B. SSO suppression effect verification

According to the parameter tuning principle, the DP is 1.25+464.78i under the current worst condition that all the units are at full power. The DP implies that SSO may occur under the current worst condition. Then the parameter pre-tuning algorithm is started up following Table I, and $k_{sso}$ is pre-tuned to 0.0144. The parameter tuning algorithm only takes 17.16ms, which is qualified for online application.

*Scenario 1:* The pre-disturbance operating condition is shown in Table II. The active power of each unit changes from 1.5MW to 2.0MW at $t$=0.2s.

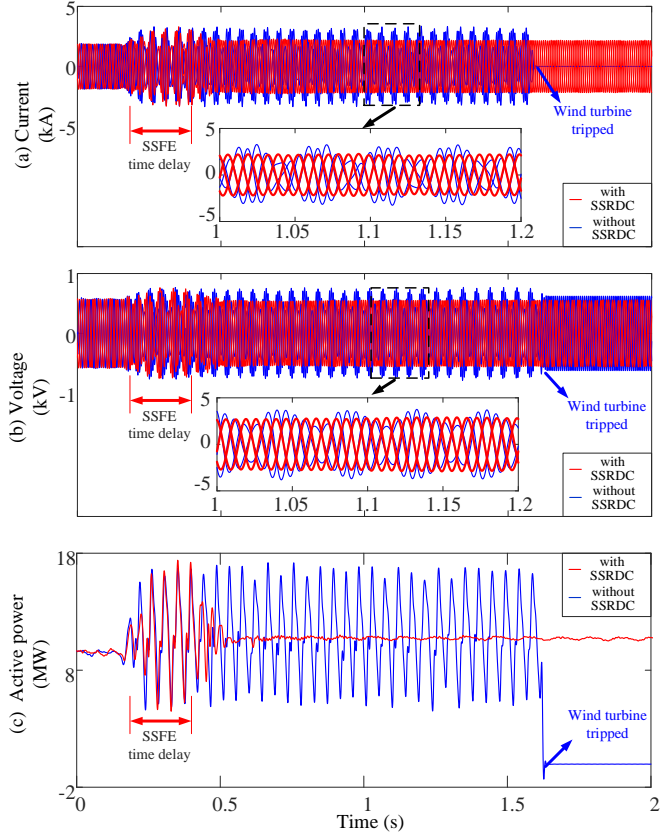

Fig. 8. Dynamics of the D-PMSG wind farm in scenario 1: (a) Three-phase current of a D-PMSG turbine. (b) Three-phase voltage of a D-PMSG turbine. (c) Active power of the D-PMSG wind farm.

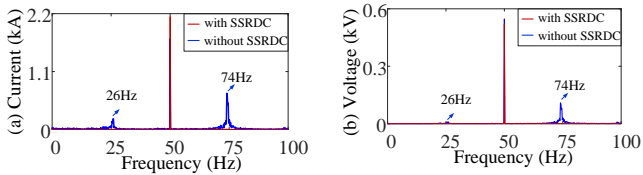

Fig. 9. Frequency domain analysis of current and voltage in scenario 1.

Scenario 1 is used to test the SSO suppression effects of the proposed SSRDC. In scenario 1, the DP is -4.87+2*π*73.39*i before 0.2s, indicating that the system is stable. However, as the active power increases to 2MW at $t$=0.2s, the DP changes to 1.25+2*π*73.97*i, then SSO occurs.

Fig. 8 shows the current, voltage, and active power of the wind farm in scenario 1. The blue lines clearly show that the D-PMSG wind farm remains stable before 0.2s, and oscillates without SSRDC after increasing the output power. Due to the severe oscillation, the relay of the D-PMSG controller activates and the wind turbines are tripped at $t$=1.6s. By comparison, the D-PMSG wind turbines with the tuned SSRDC can effectively suppress the oscillation, as shown in the red lines in Fig. 8. The response of the SSRDC is fast and effective, thus the active power oscillation in Fig. 8 (c) is damped rapidly within 100ms after the time delay of SSFE.

Fig. 9 gives the frequency spectrogram of scenario 1. For the system without SSRDC, the oscillation frequency contains a super-synchronous frequency component, 74Hz, which is close to 73.97Hz, the imaginary part of DP. The simulation results thus verify the accuracy of the transfer function's poles-based SSO analysis method. Moreover, a sub-synchronous component, 26Hz, is coupled to 74Hz in the frequency domain, which is also consistent with existing research[30]. The proposed SSRDC suppresses the SSO rapidly, thus there are nearly no sub-/super- synchronous frequency components, proving the effectiveness of the proposed SSRDC.

As the active power can directly represent the stability of the D-PMSG wind turbines, the rest of the paper will only present the active power waveform to show the dynamics of wind turbines.

## C. Adaptability test

An ideal SSRDC needs to adapt to various operating conditions, therefore, the performance of the proposed SSRDC under common/extreme condition changes is tested.

### 1) Adaptability to common condition changes

*Scenario 2:* The pre-disturbance operating condition is shown in Table II. The number of online units changes from 6 to 4, 8, and 9 at $t$=1s respectively.

Scenario 2 is used to test the adaptability of the proposed SSRDC to the common condition changes on the wind farm side. According to the tuning principle, the value of $k_{sso}$ needs to be updated in scenario 2. Table IV lists the tuned values of $k_{sso}$ and DPs before/after the changes.

TABLE IV
DP AND $K_{SSO}$ WHEN THE NUMBER OF ONLINE UNIT CHANGES

| Units changes | DP before units change | DP after units change | $k_{sso}$ | DP after $k_{sso}$ updates |
|---|---|---|---|---|
| 6→4 | -4.87+461.1i | -14.38+456.02i | 0 | -4.87+461.11i |
| 6→8 | -4.87+461.1i | 4.99+468.44i | 0.042 | -22.67+481.49i |
| 6→9 | -4.87+461.1i | 12.03+470.72i | 0.078 | -36.15+541.19i |

Table IV shows that when the number of online units decreases from 6 to 4, $k_{sso}$ is updated to zero because SSO accidents will not occur even under the worst condition. However, as the number of online units increases, the risk of SSO instability also grows accordingly. Thus, the real part of DP in Table IV became larger, indicating the worse stability of the system. Consequently, the tuned values of $k_{sso}$ also become greater to handle the worse situations. After $k_{sso}$ is updated online, the DP of the system returns to stable status. Since each unit is not at its full power, therefore, the DPs after $k_{sso}$ updates reserve a significant stable margin, which can be inferred from the real part of DP after $k_{sso}$ updates. It is noted that the time of updating $k_{sso}$ is in the range of 10-40ms, which is close to that of pre-tuning $k_{sso}$.

Fig. 10 presents the active power waveforms with/without SSRDC in scenario 2. Fig. 10(a) shows that the active power curve realizes a smooth transition from 6*1.5MW to 4*1.5MW without SSO's occurrence. Fig. 10(b) and (c) show that severe SSO occurs without SSRDC after the number of online units increases at *t*=1s. And the oscillations are successfully suppressed by the proposed SSRDC within 150ms, as shown in the red line. The simulation results are consistent with the analysis results of Table IV and verify the validation of the proposed SSRDC.

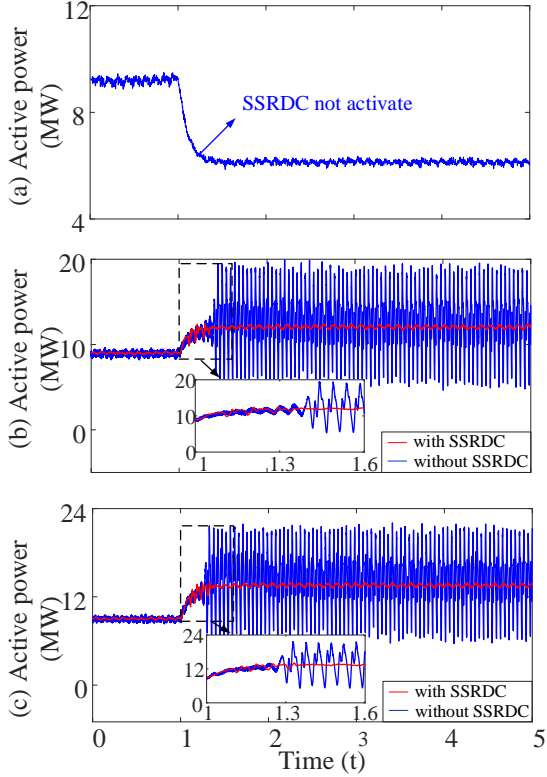

Fig. 10. Dynamics of wind farm's active power in scenario 2. The number of online units changes at *t*=1s from 6 to (a) 4. (b) 8. (c) 9.

*Scenario 3:* The pre-disturbance operating condition is shown in Table II. The grid impedance changes from $Z_L$ to 1.3$Z_L$, 1.35$Z_L$, and 1.4$Z_L$ at *t*=1s respectively.

Scenario 3 is used to test the adaptability of the proposed SSRDC to the common condition changes on the grid side. Table V lists the tuned values of $k_{sso}$ and DPs before/after the changes.

It is inferred from the real-part of DP after changes that the system stability becomes worse with the increase in gird impedance. To this end, the value of $k_{sso}$ is tuned greater accordingly. And the DP after $k_{sso}$ updates indicates that enough stability margin is reserved for scenario 3. The changing trends of DP and $k_{sso}$ that respond to the increase in grid impedance are also consistent with the findings in section II.

Fig. 11 presents the active power waveforms with/without SSRDC in scenario 3. The simulations show that severe SSO occurs without SSRDC after the gird impedance increases, which verifies the analysis of Table V. The wind farm relay activates and trips the wind turbine eventually. By comparison, the proposed SSRDC rapidly suppresses the oscillations within 150ms, as shown in the red lines. The adaptability of the proposed SSRDC to grid side impedance change is thus demonstrated.

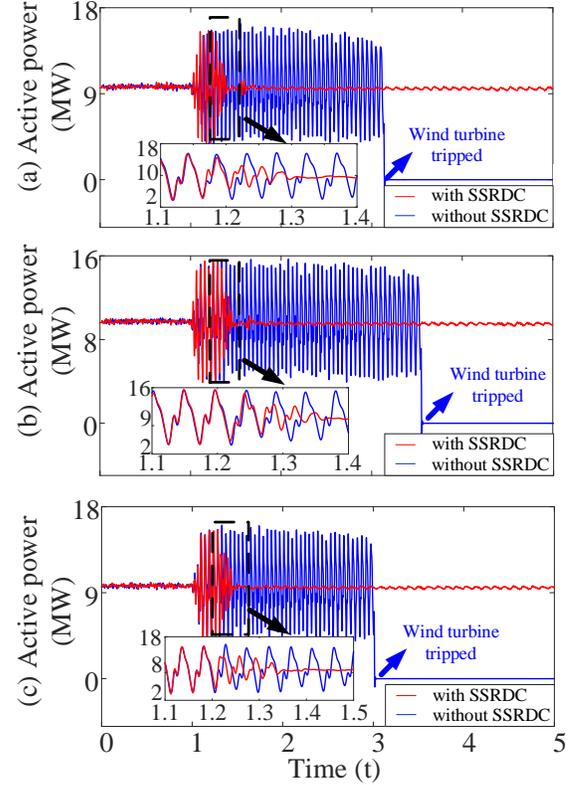

Fig. 11. Dynamics of wind farm's active power in scenario 3. The grid side impedance increases at *t*=1s from $Z_L$ to (a) 1.3$Z_L$. (b) 1.35 $Z_L$. (c) 1.4 $Z_L$.

TABLE V
DP AND $K_{SSO}$ WHEN GRID SIDE IMPEDANCE CHANGES

| Impedance changes | DP before impedance change | DP after impedance change | $k_{sso}$ | DP after $k_{sso}$ updates |
|---|---|---|---|---|
| $Z_L$→1.3$Z_L$ | -4.87+461.1i | 3.45+468.36i | 0.036 | -20.39+478.71i |
| $Z_L$→1.35$Z_L$ | -4.87+461.1i | 5.39+469.28i | 0.042 | -22.88+483.15i |
| $Z_L$→1.4 $Z_L$ | -4.87+461.1i | 7.42+470.14i | 0.048 | -26.47+489.22i |

*2) Adaptability to extreme conditions changes*

When serious accidents such as extreme weather or severe failure occur, the operating condition including the number of online units or the grid side impedance will change significantly. This section is designed to verify the adaptability of SSRDC to the above-mentioned extreme disturbances.

*Scenario 4:* In a strong wind speed scenario, only one wind turbine is connected to the grid at its rated power. Then the wind speed slows down, and the number of online units increases from 1 to 7, 8, and 9 respectively at *t*=1s. Each unit outputs at its rated power.

Scenario 4 is used to test the adaptability of the proposed SSRDC to the extreme condition changes on the wind farm side. Table VI lists the tuned values of $k_{sso}$ and DPs before/after the changes.

Similar to Table IV, the increase in online units leads to the growth of the real-part of DP and $k_{sso}$. Although the tuned $k_{sso}$ are the same as that in Table IV, the real part of the DP after $k_{sso}$ updates indicates that the stability margin in scenario 4 is much less than that in scenario 2. The reason is that scenario 4 is already under its current worst condition but scenario 2 is not yet the worst. Nevertheless, the proposed SSRDC leaves an adequate stable margin to suppress the SSO in scenario 4.

TABLE VI
DP AND $K_{SSO}$ WHEN THE NUMBER OF ONLINE UNITS EXTREMELY CHANGES

| Units changes | DP before units change | DP after units change | $k_{sso}$ | DP after $k_{sso}$ updates |
|---|---|---|---|---|
| 1→7 | -20.56+448.62i | 8.38+468.01i | 0.024 | -5.10+474.09i |
| 1→8 | -20.56+448.62i | 16.53+470.54i | 0.042 | -4.98+486.63i |
| 1→9 | -20.56+448.62i | 25.409+471.92i | 0.078 | -5.28+527.96i |

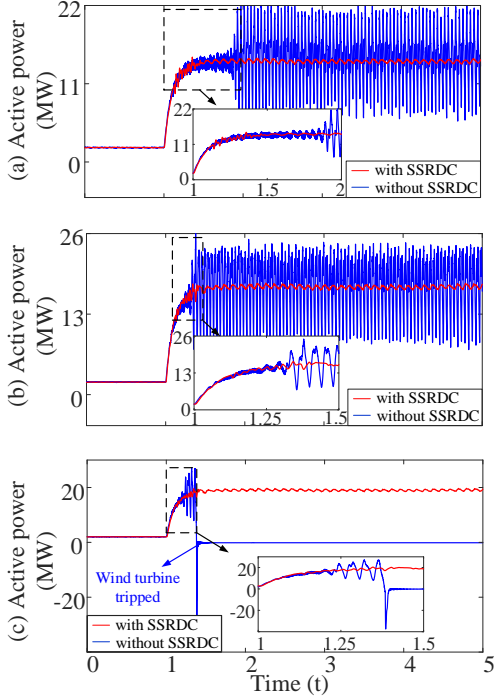

Fig. 12. Dynamics of wind farm's active power in scenario 4. The number of online D-PMSGs changes at $t$=1s from 1 to (a) 7. (b) 8. (c) 9.

Fig. 12 presents the active power waveforms with/without SSRDC in scenario 4. It shows that the oscillation occurs under extreme changes in wind farms without SSRDC. The transition from steady-state to oscillation can be obviously observed, which is related to the severity of disturbances. The larger changes in the number of online units lead to faster transitions and more severe oscillations. By comparison, the system maintains its stability with the proposed SSRDC despite that an almost invisible oscillation exists during the transitions.

*Scenario 5:* In a strong grid scenario, the grid impedance is $0.3Z_L$ before the disturbance, and six PMSG wind turbines are connected to the grid on their rated power. Then the grid impedance changes extremely from $0.3Z_L$ to $1.3Z_L$, $1.35Z_L$, and $1.4Z_L$ respectively at $t$=1s due to severe faults.

Scenario 5 is used to test the adaptability of the proposed SSRDC to the extreme condition changes on the grid side. Table VII lists the tuned values of $k_{sso}$ and DPs before/after the changes.

TABLE VII
DP AND $K_{SSO}$ WHEN GRID SIDE IMPEDANCE EXTREMELY CHANGE

| Impedance changes | DP before impedance change | DP after impedance change | $k_{sso}$ | DP after $k_{sso}$ updates |
|---|---|---|---|---|
| $0.3Z_L$→$1.3Z_L$ | -18.8+450.1i | 14.61+470.73i | 0.036 | -4.89+484.17i |
| $0.3Z_L$→$1.35Z_L$ | -18.8+450.1i | 17.17+471.48i | 0.042 | -5.26+489.56i |
| $0.3Z_L$→$1.4Z_L$ | -18.8+450.1i | 19.79+472.12i | 0.048 | -5.28+496.16i |

Compared with scenario 3, scenario 5 faces more severe stability challenges due to the significant impedance increase and larger units' output power. However, the proposed SSRDC guarantees the SSO suppression effects with updated $k_{sso}$.

Fig. 13 presents the active power waveforms with/without SSRDC in scenario 5. Without SSRDC, serious oscillation occurs immediately after the extreme changes in grid impedance. And these oscillations can be eventually suppressed by the proposed SSRDC, as shown in red lines. However, it is worth noting that the disturbance is so severe that the oscillation cannot be suppressed immediately. A period of obvious damped oscillation which lasts 400ms can be observed during the transition. Such short-term damping oscillation is allowed in an actual power grid.

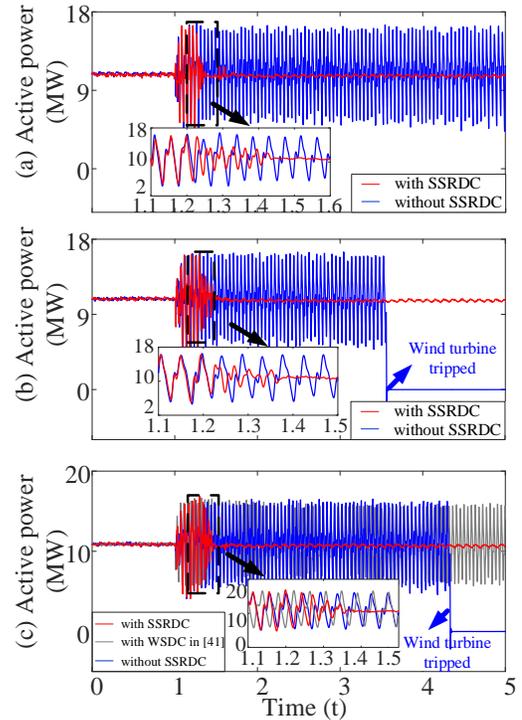

Fig. 13. Dynamics of wind farm's active power in scenario 5. The grid side impedance increases at $t$=1s from $0.3Z_L$ to (a) $1.3Z_L$. (b) $1.35 Z_L$. (c) $1.4 Z_L$

Fig. 13 (c) further compares the SSO suppression performance of the proposed SSRDC and the wideband sub-synchronous damping controller (WSDC) in [41] under extreme conditions. The damping control loop is added to the current and voltage control loop of the grid side converter in the WSDC. The result shows that the proposed SSRDC quickly suppresses the SSO within 400ms, but the oscillation continues with the WSDC. Despite that WSDC fails to suppress the oscillation eventually, the amplitude of the oscillation is smaller than that without WSDC, avoiding the tripping of wind farms. The inadaptability to the extreme condition changes of the WSDC is reasonable because the control parameters are tuned for common conditions changes. By contrast, the proposed method has better adaptability to significant operating condition changes due to the simplicity of parameter tuning and the capacity of online parameter updates.

## V. Discussion

### A. Requirement for the Impedance Model

In the proposed method, the impedance model of PMSG is used to compute the pole in (3) to tune the parameter $k_{sso}$ in (14). It should be specially stated that the proposed SSRDC doesn't depend on a specific impedance model because the stability criterion in (3) is universal. In other words, if the SSRDC is tuned with another accurate impedance model that is different from the one used in this paper, the effectiveness of the SSRDC can also be expected. The requirements for the impedance model are listed as follows.

**Accuracy**. The impedance model should be accurate, especially at the sub/super-synchronous frequency band, and the influence of PLL must be considered in the model. In this way, the parameter $k_{sso}$ of SSRDC can be optimally determined by (3). Errors in the impedance model can cause the SSRDC to be under-or over-damped.

**Simplicity**. A simple impedance model is recommended on the premise of ensuring accuracy. The complexity of the impedance model of PMSG directly determines the computation burden of solving (3).

According to these two requirements, this paper adopts a relatively simple impedance model proposed in [31]. Though the direct voltage control loop is ignored in the impedance model, it actually exists and works in the CHIL experiments. And the experimental results prove that the SSRDC tuned with such an impedance model is effective and proper. This also shows that the proposed SSRDC has good tolerance to the impedance model.

### B. Requirement for the SSFE

For the proposed SSRDC, it is vitally important to accurately and rapidly estimate the sub/super-synchronous frequency to determine the center frequency of BPF when SSO occurs. Regarding the issue of inter-harmonic frequency estimation in SSO, recent studies have provided a variety of effective methods and algorithms. Literature [42] proposes a sub-synchronous frequency component estimator to track the oscillation frequency, the time delay being 1s. The recursive DFT-based sub-synchronous frequency estimator is used in [19] with a time delay of 500ms, which proves to be acceptable for SSRDC design as SSOs may take seconds or even minutes to build up in a practical power system. The latency of the sub-synchronous frequency estimation is further reduced to 100ms by the improved iterative-TFM method in [32]. And the adaptive extend Kalman filtering approach proposed in [43] can track the time-varying sub-synchronous components within 50ms. The significant development of inter-harmonic frequency estimation technology is of great benefit for SSRDC design.

For the SSRDC proposed in this paper, the super-synchronous frequency estimation methods in [32, 43] can both meet the accuracy and latency requirements of SSFE. The oscillation frequency of SSO is assumed to be accurately obtained by SSFE. To leave sufficient time margin for frequency estimation and avoid possible maloperation of SSRDC, it is reasonable to set the time delay of SSFE as 200ms.

### C. Computation efficiency of parameter tuning

The simplicity of parameter tuning is one of the most prominent advantages of the proposed SSRDC. The only parameter to be tuned online is $k_{sso}$, whose tuning time depends on two factors: the number of iterations and the calculation time per iteration. The number of iterations is determined by the operating condition, normally the maximum number $N_{max}<30$. The calculation time of each iteration is mainly determined by the computation complexity of solving poles in (3), which is influenced by the number of wind turbines and the order of the impedance model. The following discusses the computation efficiency of the proposed method under the unified wind turbines assumption and multiple wind turbine assumptions.

The unified wind turbines assumption refers to the scenario in which all wind turbines in the wind farm have the same control parameters and operating status. It is an ideal assumption, but it has to be taken under the condition of limited experimental equipment[18]. Under such assumption, the impedance of each PMSG turbine is the same, and the total impedance of the wind farm is computed by (8). Obviously, the increase in the number of online units will hardly affect the calculation efficiency of parameter tuning.

In the multiple wind turbine assumption, the differences of PMSG turbines in control parameters and operating status are considered. Therefore, the total impedance of the wind farm is computed by (15).

$$Z_{PMSG} = [Z_{pmsg-1}(s) + X_{T1}(s)] / / ... / / [Z_{pmsg-m}(s) + X_{Tm}(s)], \quad (15)$$

where $m$ is the number of wind turbine types, $m \leq n$.

Compared with (8), the impedance in (15) has a higher-order and more complicated formula, which undoubtedly increases the computation burden of parameter tuning. Fortunately, the proposed method can greatly alleviate this problem because the online parameter tuning is pre-performed under the current worst condition. In other words, the impedance difference caused by different operating statuses is avoided, and only the difference caused by different control parameters needs to be accounted for. With the aid of the above parameter tuning principle, the number of wind turbine types can be reduced to a very small number, such as 3 to 5. Fig. 14 presents the computation efficiency of parameter tuning under the different number of wind turbine types. It shows the calculation time of each iteration increases with more wind turbine types, but the time is still less than 10ms. The maximum total time $t_{max}$ for parameter tuning is $t_{max}=\tau*N_{max}$, which is in the range of 0.18s to 0.3s. The computation time is acceptable for the SSRDC because the parameter is pre-tuned online before the SSO occurs.

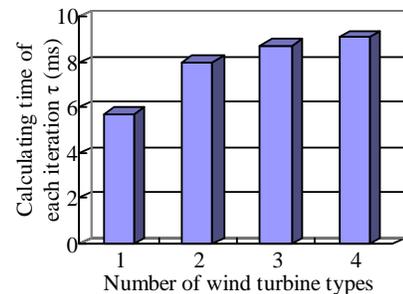

Fig. 14. Computation efficiency of parameter tuning with multiple wind turbine types

*D. Impacts on PLL's response*

The core of the proposed method is to reduce the bandwidth of the PLL when SSO occurs. Thus, one of the most concerned questions is whether the proposed SSRDC influences the normal function of PLL. A brief analysis is given below, and the conclusion is the proposed method would not negatively influence the steady-state and dynamic response of PLL.

**Impacts on PLL's steady-state response.** Under steady-state, the output of the SSFE would be $f_{sso}=0$ Hz, thus no signals could pass through the BPF. In other words, the transfer function of the SSRDC is always $H_{pll\_sso}(s)=0$ under a steady state. Therefore, the proposed SSRDC would not influence the steady-state response of PLL.

**Impacts on PLL's dynamic response.** The dynamic response of PLL refers to the ability to track the phase during the transient state, especially during the grid fault stage. Under such a dynamic process, there will be certain sub/super-synchronous components (SSCs) even if no SSO occurs, which may lead to incorrect start-up of the proposed SSRDC.

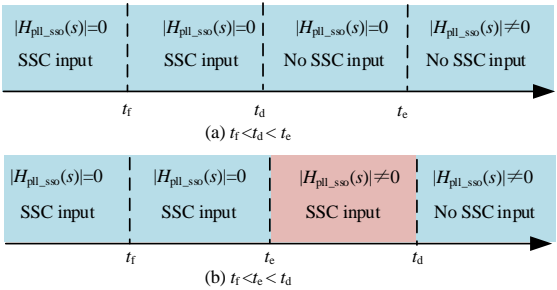

Fig. 15. The output of the PLL based SSRDC during grid fault dynamics

Denote $t_f$ as the time of fault duration, which is less than 100ms due to the operation of relay protection. Denote $t_d$ as the decay time of SSCs, usually in the range of 100ms-200ms. And $t_e$ is denoted as the time delay of SSFE, which is larger than 100ms. According to the relationship between $t_d$ and $t_e$, there are two different dynamic processes, as shown in Fig.15.

For the case of $t_f < t_d < t_e$, the SSC decays before the delay of SSFE. When $t<t_e$, the output of SSFE is 0 Hz, thus the transfer function of the proposed SSRDC $H_{pll\_sso}(s)=0$. Despite SSCs existing in the period of $t<t_d$, there's no path for them in the SSRDC. After $t_e$, the SSFE may output a certain super-synchronous frequency, thus the SSRDC incorrectly startup. But there are no SSCs after $t>t_d$, the output of SSRDC is 0. Therefore, the output of PLL-based SSRDC is always 0 in the case of $t_f<t_d<t_e$, indicating that the proposed method would not influence the dynamic response of PLL under grid faults.

For the case of $t_f <t_e <t_d$ the output of SSRDC remains 0 in the period of $t<t_e$ and $t>t_d$. During $t_e < t < t_d$, SSCs exist and the SSRDC is active, thus the proposed SSRDC may have certain outputs. However, its impacts on PLL dynamic response are almost negligible because the SSRDC only works on a very narrow sub/super-synchronous frequency band determined by the BPF. Most of the frequency components are unaffected.

To verify the correctness of the above analysis, a scenario of grid fault is designed as follows.

*Scenario 6.* A three-phase short-circuit fault occurs in the power grid near the wind farm at $t$=1.5s, and the fault is eliminated at $t$=1.55s.

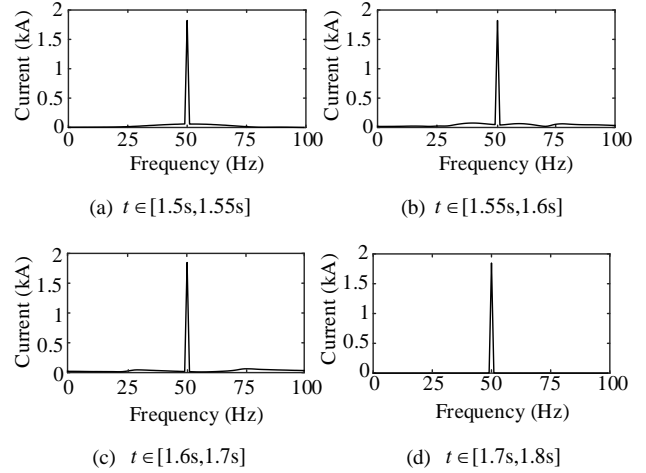

Fig.16. Frequency spectrum at different stages of grid fault

Fig. 16 presents the frequency spectrum of the wind farm's output current at different stages. Fig. 16(a) shows that the frequency spectrum at the fault stage is dominated by the fundamental component, accompanied by broad-band and low-amplitude SSCs. After the fault is eliminated, the SSCs gradually decay in Fig.16 (b) and (c), and finally decay to 0 in Fig.16(d). In this scenario, $t_d$ is in the range of 150ms to 200ms.

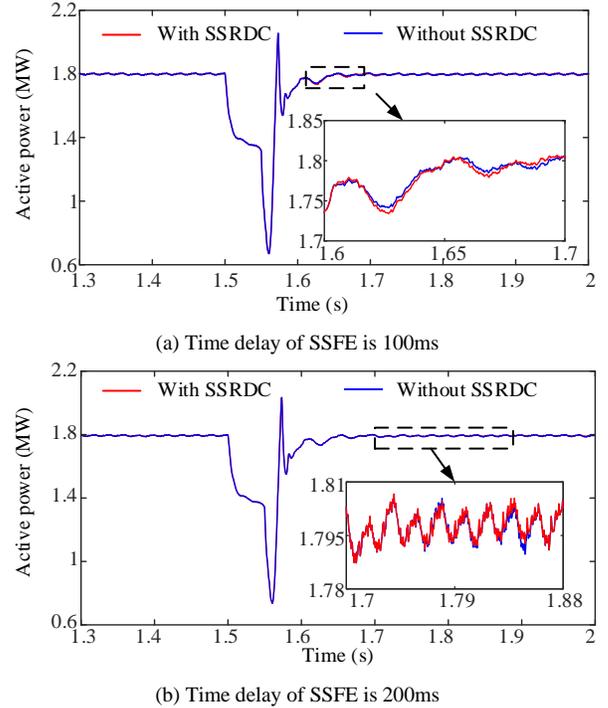

Fig.17. Impacts of SSRDC on PLL's dynamic responses under grid fault

Fig.17 shows the impacts of the proposed SSRDC on PLL's dynamic responses. In Fig.17(a), $t_e$ is 100ms, thus it belongs to the case of $t_f < t_e < t_d$, indicating the SSRDC starts up before SSCs decay to 0. The results show that although the SSRDC is activated, the PLL's dynamic performance is almost unchanged. This is because, as analyzed previously, the SSRDC only affects a very narrow band of frequency components, whereas the SSCs feature a broad frequency band and low amplitude during the grid fault process. In Fig.17(b), $t_e$ is 200ms, indicating that SSCs already decay to 0 before SSRDC starts up. The results show that the dynamic responses of PLL are exactly the same with/without the SSRDC, which confirms the conclusion that the output of SSRDC is always 0 under the case of $t_f < t_d < t_e$. Therefore, the potential impacts of the proposed SSRDC on PLL can be completely avoided by properly setting the time delay of sub/super-synchronous frequency estimation.

To conclude, the proposed SSRDC would not influence the PLL's normal function. The steady-state responses and dynamic responses of PLL are almost unchanged after adding the SSRDC.

## VI. Conclusion

In this paper, a PLL-based SSRDC is designed to suppress the SSO in D-PMSG integrated power systems. Based on the DP analysis of the impedance closed-loop transfer function, the PLL parameter is found to play an essential role in system stability. Inspired by this finding, this paper then designs a PLL-based SSRDC, which features a simple structure, easy tuning, and flexible adaptability. The simple structure of the proposed SSRDC is achieved by the avoidance of phase compensation. Only one parameter needs to be tuned following two principles. These two principles achieve a proper balance among simplicity, robustness, and adaptability. The tests on the CHIL platform prove that the proposed SSRDC can not only effectively suppress the oscillations under normal operating conditions, but also adapt to the common or even extreme operating condition changes. Analysis shows that the SSRDC has high computation efficiency and good tolerance to impedance models and frequency estimation algorithms. Most importantly, the proposed SSRDC does not influence the PLL's steady-state and dynamic response.

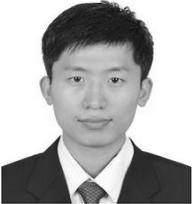

**Songhao Yang** (S'18-M'19) was born in Shandong, China, in 1989. He received the B.S. and Ph.D. degrees in electrical engineering from the Xi'an Jiaotong University, Xi'an, China, in 2012 and 2019, respectively. Besides, he received the Ph.D. degree in electrical and electronic engineering from Tokushima University, Japan, in 2019. Currently, he is an Assistant Professor at Xi'an Jiaotong University. His research interest includes power system control and protection.

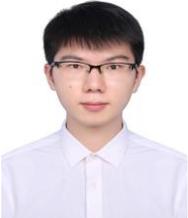

**Ruixin Shen** (S'21) is presently a postgraduate student of Xi'an Jiaotong University, Xi'an, China. He has worked on the stability analysis and control of sustainable energy integrated power systems.

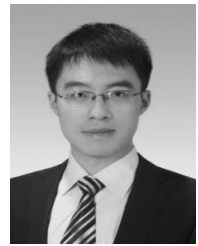

**Jin Shu.** (M'20) was born in Xi'an, China, in 1983. He received his Ph.D. degrees in electrical engineering from Xi'an Jiaotong University, Xi'an, China, in 2012. He has been a Senior Engineer with Xi'an Thermal Power Research Institute CO.LTD. His research interest includes new energy power generation.

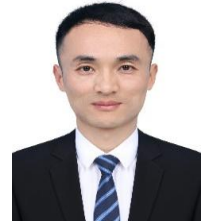

**Tao Zhang.** (S'20) received the B.S. degree in electrical engineering from Xi'an Jiaotong University, Xi'an, Chian. And he is currently pursuing a Ph.D degree. His rearch interests includes modeling and stability analysis of wind power generation systems

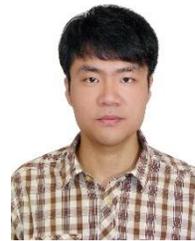

**Yujun Li.** (M'20) received the B.Sc. degree from Xi'an Jiaotong University, Xi'an, China, in 2011, the M.Sc. degree from Zhejiang University, Hangzhou, China, in 2014, and the Ph.D. degree from Hong Kong Polytechnic University, Hung Hom, Hong Kong, in 2017, all in electrical engineering. In 2017, he joined the School of Electrical Engineering, Xi'an Jiaotong University, where he was a Lecturer and is currently an Associate Professor with the School of Electrical Engineering, Xi'an Jiaotong University. His main fields of interest include grid integration of renewable energy and high-voltage directing current modeling, fault analysis and detection of power systems

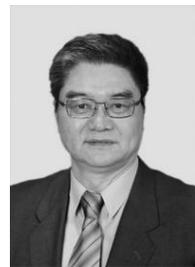

**Baohui Zhang.** (SM'99-'F'19) was born in Hebei Province, China, in 1953. He received the M.Eng. and Ph.D. degrees in electrical engineering from Xi'an Jiaotong University, Xi'an, China, in 1982 and 1988, respectively. He has been a Professor in the Electrical Engineering Department at Xi'an Jiaotong University since 1992. His research interests are system analysis, control, communication, and protection.

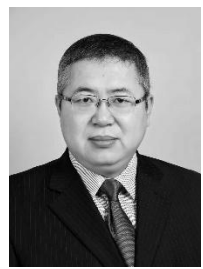

**Zhiguo Hao.** (M'10) was born in Ordos, China, in 1976. He received his B.Sc. and Ph.D. degrees in electrical engineering from Xi'an Jiaotong University, Xi'an, China, in 1998 and 2007, respectively. He has been a Professor with the Electrical Engineering Department, Xi'an Jiaotong University. His research interest includes power system protection and control.